\documentclass[a4paper]{jpconf}
\usepackage{graphicx}
\usepackage{amsmath}
\usepackage{amssymb}
\usepackage{mathrsfs}
\usepackage{extarrows}
\usepackage{fancyhdr}
\fancypagestyle{firstpage}{%
  \lhead{Journal of Physics: Conference Series, accepted (2020)
	\newline
	The proceedings of the International conference PhysicA.SPb/2020}
  \rhead{}
}

\begin{document}
\title{The Sobolev approximation for radiation transport with line overlap and continuous opacity}

\author{A V Nesterenok}

\address{Ioffe Institute, 26 Polytechnicheskaya St., 194021, Saint Petersburg, Russia}

\ead{alex-n10@yandex.ru}
\thispagestyle{firstpage}  

\begin{abstract}
The radiation transport problem in the plane--parallel medium with the large velocity gradient is considered. The Sobolev approximation is used. The effects of continuum absorption and line overlap are taken into account. The photon loss probability functions are calculated and tabulated. Two calculations are performed -- for the Gaussian spectral line profile and for the rectangular profile. It is shown that at particular choice of the rectangular profile width the results of the calculations are very close. The evaluated photon loss probability functions may be used in the calculations of energy level populations of OH molecule in the interstellar gas flows.
\end{abstract}

\section{Introduction}
The escape-probability method to the solution of radiative transfer problem in a medium with large velocity gradient was formulated for the first time by \cite{Sobolev1957}. A number of generalizations of the escape-probability approach were developed taking into account the continuum opacity \cite{Hummer1985} and the line overlap \cite{Bujarrabal1980,Elitzur1985,Doel1990,Pavlakis1996}. See also reviews \cite{Rybicki1984,Grinin1984}.

The spectral line overlap takes place when the frequencies of two or more spectral lines are close, and the thermal broadening or/and the velocity gradient presented in the interstellar gas lead to the overlapping of spectral line profiles. The OH radical is one of the most important molecules in the interstellar medium, and it is the specimen for which the phenomenon of far-infrared line overlap is essential in the formation of molecular radiation \cite{Doel1990}. In particular, the overlap of the far-infrared lines in OH may lead to the population inversion for the transitions of the lower $\Lambda$-doublet at 18~cm \cite{Litvak1969}. The rotational energy levels of OH molecule are split by spin--orbit coupling into two sets $^2\Pi_{1/2}$ and $^2\Pi_{3/2}$. Individual levels are split by the $\Lambda$-doubling and by the nuclear hyperfine interaction. The frequency shift of the certain far-infrared transitions connecting the different rotational states is of the order of few MHz, that is comparable to the thermal (microturbulent) width \cite{Burdyuzha1973}.   

Here, we use the formalism given in \cite{Hummer1985} to calculate the photon loss probability functions in the medium with high velocity gradient. We extend the calculations by \cite{Hummer1985} taking into account the effect of the line overlap.

\section{Formulation}
Consider the line formation problem in a plane--parallel medium oriented perpendicular to the axis $z$ and having a monotonic velocity field $v(z)$ along the $z$ axis. In this geometry the line intensity $I$ at frequency $\nu$ depends on the depth $z$ and on the angle $\theta$ between the axis $z$ and the given direction. We will use the parameters $\mu = cos \theta$ and dimensionless frequency shift $x$,

\begin{equation}
x = \frac{\nu - \nu_{ik}}{\Delta \nu_{D}},
\end{equation}

\noindent
where $\nu_{ik}$ is the center frequency of the spectral line, $\Delta \nu_{D}$ -- line profile width, $i$ and $k$ -- the numbers of upper and lower energy levels of the atom or molecule. We assume that complete redistribution occurs in the scattering of the photon, which implies that the emission and absorption profile functions are the same. The Gaussian spectral line profile is 

\begin{equation}
\displaystyle
\phi(x) = \frac{1}{\sqrt{\pi}}\text{exp}(-x^2).
\end{equation}

Initially we consider the radiation transport without line overlap. It is assumed that there is no inversion of level populations for the considered pair of levels. The equation of radiative transfer is

\begin{equation}
\displaystyle
\mu \frac{dI(z,\mu,x)}{dz} = \left[-\kappa_L(z) I(z,\mu,x) + \varepsilon_L(z) \right] \phi\left[x - \mu\frac{v(z)}{v_D}\right] - \kappa_C I(z,\mu,x) + \varepsilon_C,
\label{eq_radtransf1}
\end{equation}

\noindent
where $\varepsilon_C$ and $\kappa_C$ -- the continuum emissivity and opacity, $\varepsilon_L(z)$ and $\kappa_L(z)$ -- the line emissivity and opacity averaged over the line profile, respectively, $v_D$ is the line profile width in velocity units. The Doppler shift of the line frequency relative to the moving medium is taken into account in the argument of the line profile. For the rate equations for the populations of energy levels, the average radiation intensity is required. The intensity in the spectral line averaged over the line profile and direction is 

\begin{equation}
J(z) = \frac{1}{2} \int\limits_{-1}^{1} d\mu \int\limits_{-\infty}^{\infty} dx \, \phi\left[x - \mu\frac{v(z)}{v_D}\right] I(z,\mu,x). 
\end{equation}

\noindent
The essential assumption of the escape probability method (in particular, the Sobolev approximation) is that the physical parameters in the medium do not change in the region where the radiation is coupled to the medium \cite{Rybicki1984}. Let us introduce the following parameters \cite{Sobolev1957,Hummer1985}

\begin{equation}
\gamma = \frac{1}{\kappa_L v_{D}}\frac{dv}{dz}, \quad \delta = \frac{1}{\kappa_C v_{D}}\frac{dv}{dz}.
\end{equation}

\noindent
In the Sobolev, or large velocity, approximation, the average intensity in the spectral line is \cite{Hummer1985}

\begin{equation}
\displaystyle
J(z) = S_L \left[1 - 2\mathcal{P}(\delta, \gamma) \right] + S_C \left[1 - \mathcal{Q}(\delta, \gamma, \tau_{c1}) - \mathcal{Q}(\delta, \gamma, \tau_{c2})\right], \\ [10pt]
\end{equation}

\noindent
where $S_L = \varepsilon_L/\kappa_L$ is the source function for the line, and $S_C = \varepsilon_C/\kappa_C$ is the source function for the continuum, $\mathcal{P}(\delta, \gamma)$ and $\mathcal{Q}(\delta, \gamma, \tau_{c})$ are one-sided loss probability functions for the photons created by line and continuum processes, respectively, $\tau_{c1}$ and $\tau_{c2}$ are the continuum optical depths to the cloud boundaries. The formulae for $\mathcal{P}(\delta, \gamma)$ and $\mathcal{Q}(\delta, \gamma, \tau_{c})$ are given by \cite{Hummer1985,Nesterenok2016}. 

Consider two spectral lines of the same atom or molecule which have close center frequencies, $\nu_1$ and $\nu_2$. In this case, the equation of radiative transfer is

\begin{equation}
\begin{array}{c}
\displaystyle
\mu \frac{dI(z,\mu,x)}{dz} = \left[-\kappa_{L1}(z) I(z,\mu,x) + \varepsilon_{L1}(z) \right] \phi\left[x - \mu\frac{v(z)}{v_D}\right] \\ [10pt]
\displaystyle
+ \left[-\kappa_{L2}(z) I(z,\mu,x) + \varepsilon_{L2}(z) \right] \phi\left[x + \Delta x - \mu\frac{v(z)}{v_D}\right] - \kappa_C I(z,\mu,x) + \varepsilon_C,
\end{array}
\label{eq_radtransf2}
\end{equation}

\noindent
where $x = (\nu - \nu_{1})/\Delta \nu_{D}$, and $\Delta x = (\nu_{1} - \nu_{2})/\Delta \nu_{D}$ is the relative difference of the line frequencies. Here we assume that $\vert\nu_{1} - \nu_{2}\vert \ll \nu_{1}$. If $\vert \Delta x \vert \lesssim 1$, the line overlap may be called local or thermal. In this case, the significant overlap between spectral lines exists even in the absence of the velocity gradient in the medium. If $\vert \Delta x \vert > 1$, the overlap is non-local -- the radiation, emitted in one spectral line overlaps with the second spectral line emitted in the neighbouring region with shifted velocity.

The mean optical depth in the first line $\tau = \kappa_{L1}z$ is used for the cloud depth. The cloud has total thickness $T$ on the $\tau$-scale. The velocity of the gas is assumed to be equal $v(\tau) = v_{D}\gamma_1\tau$ (constant velocity gradient). The line intensity in the positive direction ($\mu > 0$) -- the solution of the equation (\ref{eq_radtransf2}) may be written

\begin{equation}
\begin{array}{c}
\displaystyle
I^{+}(x, \mu, \tau) = \frac{1}{\mu} \int\limits_0^{\tau} d\tau' \left[ S_{L1} \phi\left(x - \mu\gamma_1\tau'\right) + S_{L2}\frac{\kappa_{L2}}{\kappa_{L1}} \phi\left(x - \mu\gamma_1\tau' + \Delta x\right) + S_{C} \frac{\kappa_{C}}{\kappa_{L1}} \right] \times \\ [10pt]
\displaystyle
\times \text{exp} \left\lbrace -\frac{1}{\mu} \int\limits_{\tau'}^{\tau} dt \left[\phi\left(x - \mu\gamma_1 t\right) +  \frac{\kappa_{L2}}{\kappa_{L1}} \phi\left(x - \mu\gamma_1 t + \Delta x\right) + \frac{\kappa_{C}}{\kappa_{L1}} \right] \right\rbrace
\end{array}
\label{eq_intensity+}
\end{equation}

\noindent
For the line intensity in the negative direction ($\mu < 0$)

\begin{equation}
\begin{array}{c}
\displaystyle
I^{-}(x, \mu, \tau) = \frac{1}{\vert\mu\vert} \int\limits_{\tau}^{T} d\tau' \left[ S_{L1} \phi\left(x + \vert\mu\vert \gamma_1\tau'\right) + S_{L2}\frac{\kappa_{L2}}{\kappa_{L1}} \phi\left(x + \vert\mu\vert\gamma_1\tau' + \Delta x\right) + S_{C} \frac{\kappa_{C}}{\kappa_{L1}} \right] \times \\ [10pt]
\displaystyle
\times \text{exp} \left\lbrace -\frac{1}{\vert\mu\vert} \int\limits_{\tau}^{\tau'} dt \left[\phi\left(x + \vert\mu\vert\gamma_1 t\right) +  \frac{\kappa_{L2}}{\kappa_{L1}} \phi\left(x + \vert\mu\vert\gamma_1 t + \Delta x\right) + \frac{\kappa_{C}}{\kappa_{L1}} \right] \right\rbrace
\end{array}
\label{eq_intensity-}
\end{equation}

\noindent
After the replacement $x \to -x$, $\mu \to -\mu$ and taking into account $\phi(x)=\phi(-x)$, the equation (\ref{eq_intensity-}) becomes similar to the equation (\ref{eq_intensity+}). The intensity averaged over the direction and the spectral profile of the first line is

\begin{equation}
J_{1}(\tau) = \frac{1}{2} \int\limits_{0}^{1} d\mu \int\limits_{-\infty}^{\infty} dx \, \phi\left(x - \mu\gamma_1\tau\right) \left[I^{+}(x, \mu, \tau) + I^{-}(-x, \mu, \tau) \right].
\label{eq_av_int2}
\end{equation} 

\noindent
Substituting equations (\ref{eq_intensity+}) and (\ref{eq_intensity-}) into the equation (\ref{eq_av_int2}), one can deduce

\begin{equation}
\displaystyle
J_{1}(\tau) = \int\limits_{0}^{T} d\tau' \mathcal{K}_{11}\left( \vert\tau - \tau'\vert\right) S_{L1} +  \int\limits_{0}^{T} d\tau' \mathcal{K}_{12} \left(\vert\tau - \tau'\vert\right) S_{L2} + \int\limits_{0}^{T} d\tau' \mathcal{L}_{1}\left(\vert\tau - \tau'\vert\right) S_{C},
\label{eq_av_int3}
\end{equation}
where 

\begin{equation}
\begin{array}{c}
\displaystyle
\mathcal{K}_{11}(\tau) = \frac{1}{2} \int\limits_0^1 \frac{d\mu}{\mu} \int\limits_{-\infty}^{\infty} dx \, \phi(x - \mu\gamma_1\tau) \phi(x) \times \\[10pt] 
\displaystyle
\times \text{exp} \left\lbrace -\frac{1}{\mu} \int\limits_{0}^{\tau} dt \left[ \phi(x - \mu\gamma_1 t) + \frac{\gamma_1}{\gamma_2} \phi(x - \mu\gamma_1 t + \Delta x) + \frac{\gamma_1}{\delta} \right] \right\rbrace,
\end{array}
\label{eq_kint}
\end{equation}

\noindent
and analogous expressions for $\mathcal{K}_{12}(\tau)$ and $\mathcal{L}_{1}(\tau)$, see also equations (2.12) and (2.13) in \cite{Hummer1985}. The parameters $\gamma_1$, $\gamma_2$ and $\delta$ may be either all positive or all negative depending on the sign of the velocity gradient. 

In the large velocity gradient approximation, the source functions can be brought outside the integral in the equation (\ref{eq_av_int3}). The average line intensity takes the form

\begin{equation}
\begin{array}{c}
\displaystyle
J_{1}(\tau) = S_{L1} \left[1 - \mathcal{P}_{11}(\tau) - \mathcal{P}_{11}(T - \tau) \right] + S_{L2} \left[1 - \mathcal{P}_{12}(\tau) - \mathcal{P}_{12}(T - \tau) \right] + \\ [10pt]
\displaystyle
+ S_{C} \left[1 - \mathcal{Q}_{1}(\tau) - \mathcal{Q}_{1}(T - \tau) \right],
\end{array}
\label{eq_av_int4}
\end{equation} 

\noindent
where $\mathcal{P}_{11}$, $\mathcal{P}_{12}$ and $\mathcal{Q}_{1}$ are one-sided loss probability functions. After the change of the variables $x' = x - \mu\gamma_1\tau'$ and $u = x - \mu\gamma_1 t$ in the equation (\ref{eq_kint}), the following expressions are valid for the loss probability functions

\begin{equation}
\begin{array}{c}
\displaystyle
\mathcal{P}_{11}(\tau) = \frac{1}{2} - \int\limits_{0}^{\tau} d\tau' \mathcal{K}_{11}(\tau') =  \frac{1}{2} - \frac{1}{2} \int\limits_0^1 \frac{d\mu}{\mu^2\vert\gamma_1\vert} \int\limits_{-\infty}^{\infty} dx \, \phi(x)\int\limits_{x}^{x + \mu\vert\gamma_1\vert\tau} dx' \phi(x') \times \\ [10pt] 
\displaystyle
\times \text{exp} \left\lbrace -\frac{1}{\mu^2\vert\gamma_1\vert} \int\limits_{x}^{x'} du \left[ \phi(u) + \frac{\gamma_1}{\gamma_2} \phi(u \mp \Delta x) + \frac{\gamma_1}{\delta} \right] \right\rbrace, \\ [10pt]
\displaystyle
\mathcal{P}_{12}(\tau) = \frac{1}{2} - \int\limits_{0}^{\tau} d\tau' \mathcal{K}_{12}(\tau') = \frac{1}{2} - \frac{1}{2} \int\limits_0^1 \frac{d\mu}{\mu^2\vert\gamma_2\vert} \int\limits_{-\infty}^{\infty} dx \, \phi(x \mp \Delta x) \int\limits_{x}^{x + \mu\vert\gamma_1\vert\tau} dx' \phi(x') \times \\[10pt] 
\displaystyle
\times \text{exp} \left\lbrace -\frac{1}{\mu^2\vert\gamma_1\vert} \int\limits_{x}^{x'} du \left[ \phi(u) + \frac{\gamma_1}{\gamma_2} \phi(u \mp \Delta x) + \frac{\gamma_1}{\delta} \right] \right\rbrace, \\ [10pt]
\displaystyle
\mathcal{Q}_{1}(\tau) = \frac{1}{2} - \int\limits_{0}^{\tau} d\tau' \mathcal{L}_{1}(\tau') = \frac{1}{2} - \frac{1}{2} \int\limits_0^1 \frac{d\mu}{\mu^2\vert\delta\vert} \int\limits_{-\infty}^{\infty} dx \int\limits_{x}^{x + \mu\vert\gamma_1\vert\tau} dx' \phi(x') \times \\[10pt] 
\displaystyle
\times \text{exp} \left\lbrace -\frac{1}{\mu^2\vert\gamma_1\vert} \int\limits_{x}^{x'} du \left[ \phi(u) + \frac{\gamma_1}{\gamma_2} \phi(u \mp \Delta x) + \frac{\gamma_1}{\delta} \right] \right\rbrace,
\end{array}
\label{eq_loss_prob_func}
\end{equation}

\noindent
where the upper sign '$-$' in the argument of the line profile is for the positive velocity gradient and the lower sign '$+$' is for the negative velocity gradient. The equation (\ref{eq_av_int4}) differs from the analogous equation in the case of no line overlap in the second term containing $S_{L2}$. In the limit $\gamma_2 \to \infty$, the function $\mathcal{P}_{12} \to 1/2$, and the functions $\mathcal{P}_{11}$ and $\mathcal{Q}_1$ equal to the one-sided loss probability functions in the case of no line overlap, see equations (2.15) and (2.16) in \cite{Hummer1985}.

Analogous to the case of radiative transfer of one spectral line, the following equation is valid \cite{Hummer1985}

\begin{equation}
\begin{array}{c}
\displaystyle
\mathcal{P}_{11}(\tau) + \mathcal{P}_{12}(\tau) + \mathcal{Q}_{1}(\tau) = \\ [10 pt]
\displaystyle
= 1 + \frac{1}{2} \int\limits_0^1 d\mu \int\limits_{-\infty}^{\infty} dx' \, \phi(x') \, \text{exp} \left\lbrace -\frac{1}{\mu^2\vert\gamma_1\vert} \int\limits_{x'}^{x' + \mu\vert\gamma_1\vert\tau} du \left[ \phi(u)+ \frac{\gamma_{1}}{\gamma_{2}} \phi(u \pm \Delta x) + \frac{\gamma_1}{\delta} \right] \right\rbrace \\ [20pt]
\displaystyle
\xlongrightarrow{\vert \gamma \vert \tau \to \infty} 1 + \frac{1}{2} \int\limits_0^1 d\mu \, e^{-\tau_c/\mu} \int\limits_{-\infty}^{\infty} dx' \, \phi(x') \, \text{exp} \left\lbrace -\frac{1}{\mu^2\vert\gamma_1\vert} \int\limits_{x'}^{\infty} du \left[ \phi(u)+ \frac{\gamma_{1}}{\gamma_{2}} \phi(u \pm \Delta x) \right] \right\rbrace,
\end{array}
\label{eq_loss_prob_id}
\end{equation}

\noindent
where the upper sign in the argument of the line profile is for the positive velocity gradient. In the large velocity gradient limit, $\vert \gamma \vert \tau \to \infty$, the functions $\mathcal{P}_{11}$ and $\mathcal{P}_{12}$ depend on four parameters $\delta$, $\gamma_1$, $\gamma_2$ and $\Delta x$. However, the loss probability in continuum $\mathcal{Q}_{1}$ depends on the cloud thickness and can be evaluated with the help of the equation (\ref{eq_loss_prob_id}). As a result, the average intensity in the first spectral line in the limit of the large velocity gradient is

\begin{equation}
\begin{array}{c}
\displaystyle
J_{1}(z) = S_{L1} \left[1 - 2\mathcal{P}_{11}(\delta, \gamma_{1}, \gamma_{2}, \Delta x) \right] + S_{L2} \left[1 - 2\mathcal{P}_{12}(\delta, \gamma_{1}, \gamma_{2}, \Delta x) \right] + \\ [10pt]
\displaystyle
+ S_{C} \left[ 1 - \mathcal{Q}_1 \left(\delta, \gamma_{1}, \gamma_{2}, \Delta x, \tau_{c1}\right) - \mathcal{Q}_1(\delta, \gamma_{1}, \gamma_{2}, \Delta x, \tau_{c2}) \right],
\end{array}
\label{eq_line_intens2}
\end{equation}

\noindent
In the calculations of the average intensity of the second spectral line $J_2$, the equation (\ref{eq_line_intens2}) must be used after the change $S_{L1} \leftrightarrow S_{L2}$, $\gamma_1 \leftrightarrow \gamma_2$ and $\Delta x \leftrightarrow -\Delta x$. 

The calculation of the integrals (\ref{eq_loss_prob_func}) is time-consuming process, and the loss probability functions have to be tabulated in order to employ this formalism in the calculations of the energy level populations of a specimen. The one-sided loss probability functions $\mathcal{P}_{11}$ and $\mathcal{P}_{12}$ were calculated for the grid of parameters $\gamma_1$, $\gamma_2/\gamma_1$, $\Delta x$ and $\delta$. The ranges of these parameters are taken to be equal: $\gamma_1$ -- $[10^{-6},10^{6}]$, $\gamma_2/\gamma_1$ -- $[10^{-2},10^{2}]$, $\Delta x$ -- $[-4,4]$, $\delta$ -- $[10^{2},10^{6}]$. We used the integration algorithms published in the book by \cite{Press2007}. The relative accuracy of the integration was set to be equal to 10$^{-5}$. 

\begin{figure}[h]
\centering
\includegraphics[width = 1.0\textwidth]{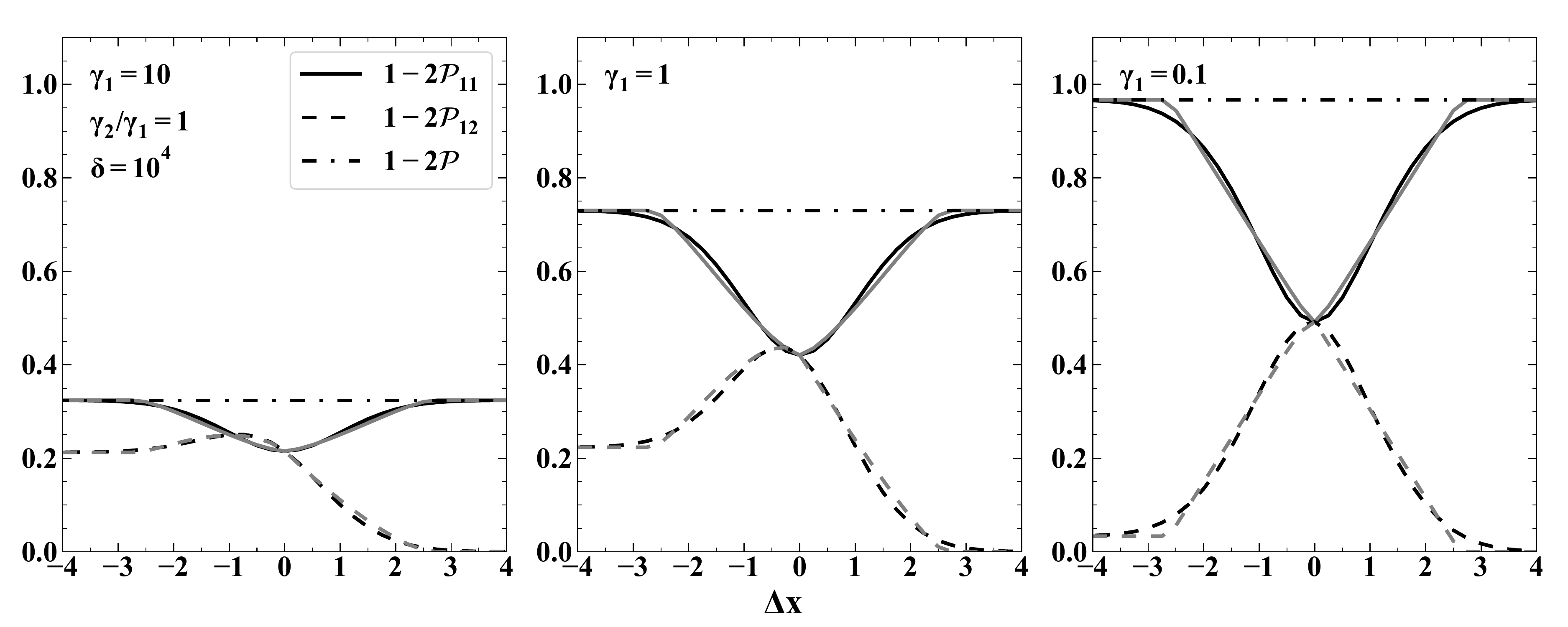}
\caption{The loss probability functions for photons created in the line processes are shown depending on the parameter $\Delta x$. The black lines correspond to the calculations where Gaussian spectral line profile is employed, the grey lines -- for the calculations with the rectangular line profile. Three cases are considered: $\gamma_1 = 10$ (optically thin case), $\gamma_1 = 1$, and $\gamma_1 = 0.1$ (optically thick case). Other parameters are fixed, $\gamma_2 = \gamma_1$, $\delta = 10^4$. The values of the function $\mathcal{P}(\delta, \gamma)$ are also presented (no line overlap).}
\label{fig_loss_functions}
\end{figure}

\section{Results}
Figure~\ref{fig_loss_functions} shows the results of the calculations of the one-sided loss probability functions $\mathcal{P}_{11}$ and $\mathcal{P}_{12}$ depending on the frequency shift between spectral lines. The results are presented for three values of the parameter $\gamma_1$, other parameters being $\gamma_2 = \gamma_1$, $\delta = 10^4$. At positive $\Delta x \geq 3$, the function $\mathcal{P}_{11}$ approaches the loss probability in the absence of the line overlap, while $\mathcal{P}_{12}$ tends to zero -- the effect of the second spectral line to the average intensity in the first line is negligible at high positive $\Delta x$ and positive velocity gradient. The function $\mathcal{P}_{12}$ is not symmetric and has a finite limit at $\Delta x \to -\infty$. The asymmetry is pronounced in the optically thin case. The radiation in the second line (that has higher frequency in this case) emitted in the neighbouring regions, is brought into the resonance with the first line whatever the frequency difference is. However, the larger $\vert \Delta x \vert$, the larger the distance between the regions where the overlapping lines are born. At high spatial separations, the assumption of the constant physical parameters (that is used in the derivation of the loss probability functions) may be no longer valid. 

Two calculations were performed -- for the Gaussian spectral line profile and for the rectangular profile, see figure~\ref{fig_loss_functions}. The width of the rectangular profile is taken to be equal to $1.5\sqrt{\pi} \Delta \nu_{D}$. The results of two simulations are close at the parameters in question. The rectangular profile can thus be used in the escape probability calculations involving the line overlap. But appropriate profile width must be chosen, see also discussion in \cite{Lockett1989}.

The typical velocity gradients in the cooling gas flow behind the C-type shock where intense OH emission arises are $10^{-11} - 10^{-10}$~cm~s$^{-1}$~cm$^{-1}$, at preshock gas densities $n_{\rm H_2} \simeq 10^{4}$~cm$^{-3}$ and cosmic ray ionization rates $\zeta \simeq 10^{-15}$~s$^{-1}$ \cite{Nesterenok2018,Nesterenok2019,Nesterenok2020}. The parameter $\Delta x \simeq 2.5$ for the closest pair of transitions joining the rotational states $^2\Pi_{1/2}$ $J = 1/2$ and $^2\Pi_{3/2}$ $J = 3/2$ of OH molecule, at $v_D \simeq 0.5$~km~s$^{-1}$. Assuming OH concentration $n_{\rm OH} \simeq 1$~cm$^{-3}$, the parameter $\vert \gamma \vert \sim 1$ for these transitions. The value of the parameter $\vert \delta \vert$ is of the order of $10^3 - 10^4$ at the far-infrared wavelengths. Thus, the photon escape due to absorption by dust is negligible at the parameters in question. If the dust temperature is much less than the gas temperature, the dust emission -- the last term in the equation (\ref{eq_line_intens2}) -- can be neglected.

\section{Conclusions}
The generalization of the Sobolev approximation for the radiation transport problem is presented for two spectral lines that may overlap locally or non-locally. The one-sided loss probability functions $\mathcal{P}_{11}$ and $\mathcal{P}_{12}$ obtained in this work may be used in the calculations of energy level populations of OH molecule in the interstellar gas flows. However, the method used in this work has a disadvantage in its extension to the problem of the overlap of more than two lines -- in this case, each of the loss probability functions will depend on too many parameters and it will be difficult to tabulate these functions.

\section*{References}
\bibliography{references} 

\end{document}